\documentstyle[aps,prl,preprint,12pt,cite]{revtex}

\begin{document}

\begin{titlepage}

\begin{flushright}
OUTP-99-28P\\
hep-th/990699\\
\end{flushright}

\begin{center}
{\Large \bf On direct and crossed channel asymptotics of four-point functions in AdS/CFT correspondence}
\vspace{0.1in}

\vspace{0.5in}
{\large  Sanjay}
\footnote{s.sanjay1@physics.ox.ac.uk}\\
{\small
Theoretical Physics,\\ 
University of Oxford, Oxford OX1 3NP, UK.}\\

\vspace{0.5in}

{\bfseries Abstract}\\

\end{center}
\vspace{.2in}
We analyse the leading logarithmic singularities in direct and crossed channel limit of the four-point functions in dilaton-axion sector of type IIB supergravity on $AdS_{5}$ in AdS/CFT correspondence. Logarithms do not cancel in the full correlator in both channels.

\vfill

\end{titlepage}

\newpage
The conjectured duality between type IIB string theory on $AdS_{5} \times S^{5}$ and $\mathcal{N}$=4 superconformal Super Yang-Mills (SYM) on the boundary of $AdS_{5}$ with gauge group SU(N) by Maldacena \cite{ma} was a subject of enormous activity in last one and half year (for a recent review see Ref. \cite{agmoo}). In the large N limit with $g_{YM}^{2}N$ fixed but large, string theory is weakly coupled and can be approximated by type IIB supergravity. Supergravity partition function on AdS can be used to compute correlation functions of operators of boundary CFT in this limit. String theoretic loop correctons correpond to $o(1/N^{2})$ corrections to correlation functions in SYM. A precise recipe to compute this is given in \cite{w,gkp}. Let $\phi_{0}^{i}(\vec{x})$ denotes the boundary value of the field $\phi^{i}(\vec{x},z)$ then the field $\leftrightarrow$ operator correspondence is $<\exp(\int d^{4}x \phi_{0}^{i}(\vec{x}) O_{i}(\vec{x}))>_{CFT}=Z_{\text{sugra}}[\phi^{i}(\vec{x}, z)|_{z=0}=\phi_{0}^{i}(\vec{x})]$, where we have denoted boundary points by a vector and z is the radial co-ordinate in AdS and $O_{i}(\vec{x})$ is the conformal operator corresponding to the bulk field $\phi^{i}(\vec{x})$. Boundary values of the fields couple to local operators in boundary CFT which acts as a source for the former. Vaccum to vaccum  amplitudes in boundary CFT can be obtained from supergravity partition function by taking functional derivatives with respect to field $\phi_{0}^{i}(\vec{x})$ and setting it to $\phi_{0}^{i}(\vec{x})=0$ after differentiation. 

Two and three point functions \cite{w,gkp,mw,fmmr,cnss,lmrs,hfs} are uniquely determined by the conformal symmetry upto an overall multiplicative factor. Four-point (and higher) functions \cite{lt,fmmr1,hf,hf1,liu,bg,bgr,bgkr,dhkmv,gr,in,cs,rps,cs1,hfmmr} contain more dynamical information then two and three point functions. Four-point functions are found to have logarithms of the conformally invariant cross ratios as two boundary points aproach each other. The correlator $<O_{\phi}(\vec x_{1})O_{C}(\vec x_{2})O_{\phi}(\vec x_{3})O_{C}(\vec x_{4})>$, where $O_{\phi}\sim Tr(F^{2} + ...), O_{C}\sim Tr(F\tilde{F} + ...)$ in the dilaton-axion sector of type IIB supergravity is evaluated in Ref. \cite{hfmmr}, direct channel (t-channel) limit ($|x_{13}|\ll |x_{12}|, |x_{24}|\ll |x_{12}|$), where $x_{12}=|\vec x_{1} - \vec  x_{2}|$ etc. , is considered and logarithms do not cancel in the full correlator in this channel. 

The leading logarithmic term is of the form  $\frac{1}{(x_{12}^{2})^{4}(x_{34}^{2})^{4}}\ln(\frac{x_{13}x_{24}}{x_{12}x_{34}})$. The leading contribution comes from an operator of dimension eight. It has been suggested \cite{agmoo,hfmmr} (following Witten) that logarithams can be interpreted as $o(1/N^{2})$ correction to the dimension of non-chiral double trace operators $:O_{\phi}O_{\phi}:$ and $:O_{C}O_{C}:$ which are dual to two-particle states in IIB supergravity. 

Another interpretation stems from conjecture \cite{kogan} that $\mathcal{N}$=4 Super Yang-Mills is a  logarithmic conformal field theory (LCFT). In such theories there are logarithms of cross ratios in the four-point functions and logarithmis are related to degenrate logarithmic operators \cite{gu}. Logrithmic operators have special two-point correlation functions \cite{gu,ckt}, with a structure that has natural interpretation in AdS \cite{kogan}. This conjecture  is further supported in \cite{ml} for the case of $AdS_{3}/CFT_{2}$ correspondence (see also \cite{gka} for a discussion of AdS/LCFT correspondence). If this is the case, then Logarithmic Conformal Field Theories  are the natural framework for AdS/CFT correspondence. Let's note that in $N\rightarrow \infty$ limit $\mathcal {N}$=4 is definately a LCFT, as the double trace operators become degenerate with chiral primaries of dimension eight. It is still an open question whether $\mathcal{N}$=4 is a logarithmic conformal field theory for finite N.

The aim of this paper is to analyse the  logarithmic singularity in crossed channels (s-channel $|x_{12}|\ll |x_{13}|, |x_{34}|\ll |x_{13}|$ and u-channel $|x_{14}|\ll |x_{13}|, |x_{23}|\ll |x_{13}|$) limits of the full correlator . Our result is that the leading logarithmic singularity in crossed channel also does not cancel and the leading contribution comes from an operator of dimension eight. We shall consider the quartic graph (defined below) and its direct and crossed channel limits as the full correlator can be written as the sum of quartic graphs only with appropriate kinemetical multiplicative factors \cite{hfmmr}.

Consider the quartic graph $D_{\Delta_{1}\Delta_{3}\Delta_{2}\Delta_{4}}(\vec x_{1}, \vec x_{3}, \vec x_{2}, \vec x_{4})$ for contact interactions in $AdS_{d+1}$ corresponding to scattering of conformal operators in boundary CFT with arbitrary dimensions $\Delta_{i}$

\begin{equation}
D_{\Delta_{1} \Delta_{3} \Delta_{2} \Delta_{4}}(\vec x_{1}, \vec x_{3}, \vec x_{2}, \vec x_{4})=\int \frac{d^{d+1}z}{{z_{0}}^{d+1}} K_{{\Delta}_{1}}(z,\vec x_{1})K_{{\Delta}_{3}}(z,\vec x_{3})K_{{\Delta}_{2}}(z,\vec x_{2})K_{{\Delta}_{4}}(z,\vec x_{4}),
\end{equation}
where $K_{\Delta}(z,\vec x)$                
\[
K_{\Delta}(z,\vec x)=(\frac{z_{0}}{z_{0}^2+(\vec{z}-\vec{x})^2})^{\Delta},
\]
is boundary to bulk propagator.

Below we consider the direct ($|x_{13}|\ll |x_{12}|, |x_{24}|\ll |x_{12}|$) and crossed ($|x_{12|}\ll |x_{13}|, |x_{34}|\ll |x_{13}|$) channel limit of the quartic graph of the form $D_{pp\Delta\Delta}(\vec x_{1}, \vec x_{3}, \vec x_{2}, \vec x_{4})$.\\
$D_{pp\Delta\Delta}$ can be written as (see equations. (A.3) and (6.3) of Ref. \cite{hfmmr})

\begin{equation}
D_{pp\Delta\Delta}=\frac{1}{x_{13}^{2p}}\frac{1}{x_{14}^{2\Delta}}\frac{1}{x_{12}^{2\Delta}}\frac{\pi^{\frac{d}{2}}}{2} \frac{\Gamma(p+\Delta-\frac{d}{2})}{\Gamma(p)\Gamma(\Delta)}\int_{0}^{\infty}du\int_{0}^{\infty}dv \frac{u^{p-1}v^{p-1}}{(u+v+uv)^{p}}\frac{1}{[(x-y)^2+uy^{2}+vx^{2}]^{\Delta}}.
\end{equation}
Defining a change of variables $\tau=\frac{uv}{u+v+uv};\; \lambda=\frac{v-u}{v+u}$, (as in Ref. \cite{hf}) we can transform $D_{pp\Delta\Delta}$ in the following form

\begin{equation}
D_{pp\Delta\Delta}=\frac{\pi^{\frac{d}{2}}}{2}\frac{\Gamma(p+\Delta-\frac{d}{2})}{\Gamma(p)\Gamma(\Delta)}\frac{2\times s^{\Delta}}{(x_{13}^{2})^{p}(x_{24}^{2})^{\Delta}}\int_{0}^{1}\frac{d\tau}{\tau}\tau^{p}\int_{-1}^{1}d\lambda\frac{(1-\tau)^{\Delta-1}(1-\lambda^{2})^{\Delta-1}}{[\tau(1-\lambda t)+s(1-\lambda^2)(1-\tau)]^{\Delta}},
\end{equation}
where 
\[
s=\frac{1}{2}\frac{(x-y)^{2}}{x^{2}+y^{2}}=\frac{1}{2}\frac{x_{13}^{2}x_{24}^{2}}{x_{12}^{2}x_{34}^{2}+x_{14}^{2}x_{23}^{2}} 
\]
and
\[
t=\frac{x^{2}-y^{2}}{x^{2}+y^{2}}=\frac{x_{12}^{2}x_{34}^{2} - x_{14}^{2}x_{23}^{2}}{x_{12}^{2}x_{34}^{2} + x_{14}^{2}x_{23}^{2}}
\]
are conformally invariant variables.

Direct (t-channel) limit $(|x_{13}|\ll |x_{12}|, |x_{24}|\ll |x_{12}|)$ corresponds to $s,t\rightarrow 0$.\\
Consider the integral
\begin{equation}
\int_{-1}^{1}d\lambda (1-\lambda^{2})^{\Delta-1}\int_{0}^{1}d\tau\frac{\tau^{p-1}(1-\tau)^{\Delta-1}}{[\tau(1-\lambda t)+s(1-\lambda^2)(1-\tau)]^{\Delta}}.
\end{equation}
As $s,t\rightarrow 0$ it can be approximated as

\begin{equation}
\sim\int_{-1}^{1}d\lambda (1-\lambda^{2})^{\Delta-1}\int_{0}^{1}d\tau\frac{\tau^{p-1}(1-\tau)^{\Delta-1}}{[\tau+s(1-\lambda^2)]^{\Delta}}.
\end{equation}
$\tau $ -integral is a hypergeometric function and we have

\begin{equation}
D_{pp\Delta\Delta}\sim\frac{\Gamma(p)\Gamma(\Delta)}{\Gamma(p+\Delta)}\int_{-1}^{1}d\lambda (1-\lambda^{2})^{\Delta-1}(s(1-\lambda^2))^{-\Delta}{_{2}F_{1}}[p, \Delta; p+\Delta;\frac{-1}{s(1-\lambda^2)}].
\end{equation}
Now the hypergeometric function $_{2}F_{1}[a, b; c; z]$ for asymptotically large values can be defined by analytic continuation and has logrithams if $(a-b)$ is an integer \cite{er,ww}. We simlply take the result from Ref. \cite{er}.

Finally carrying out the $\lambda$ integral, we obtain the leading logrithmic singularity in direct channel
\begin{equation}
D_{pp\Delta\Delta}|_{\log}=\frac{\pi^{\frac{d}{2}}}{2}(-1)^{\Delta-p-1}\frac{\Gamma(p+\Delta-2)}{\Gamma(\Delta-p+1)}\frac{2}{(\Gamma(p))^{2}}\frac{s^{\Delta}\ln s}{(x_{13}^{2})^{p}(x_{24}^{2})^{\Delta}}(\sqrt{\pi}\frac{\Gamma(\Delta)}{\Gamma(\Delta+\frac{1}{2})}),
\end{equation}
which agrees with the leading logarithmic term in equations (6.23) and (A.3) of Ref. \cite{hfmmr} including the numerical co-efficient. 

Crossed (s-channel) limit $(|x_{12}|\ll |x_{13}|, |x_{34}|\ll |x_{13}|)$ corresponds to $s\rightarrow\frac{1}{2}$ and $t\rightarrow{-1}$.\\ 
Consider the integral

\begin{equation}
I=\int_{0}^{1}\frac{d\tau}{\tau}\tau^{p}\int_{-1}^{1}d\lambda\frac{(1-\tau)^{\Delta-1}(1-\lambda^{2})^{\Delta-1}}{[\tau(1-\lambda t)+s(1-\lambda^2)(1-\tau)]^{\Delta}},
\end{equation}
which can be written as

\begin{equation}
I=\frac{(-1)^{\Delta-1}}{\Gamma(\Delta)}(\frac{\partial}{\partial s})^{\Delta-1}\int_{0}^{1}\frac{d\tau}{\tau}\tau^{p}\int_{-1}^{1}d\lambda\frac{1}{[\tau(1-\lambda t)+s(1-\lambda^2)(1-\tau)]}.
\end{equation}
Logrithmic part of integral over $\lambda$ is (eqn. 4.7b of Ref. \cite{hf})
\[
\frac{-\ln(1-t^{2})}{\sqrt{\omega^{2}-(1-t^{2})\tau^{2}}},
\]
where $\omega=2s(1-\tau)+\tau$ and $\omega\sim 1 $ in s-channel limit.

\begin{equation}
I_{\log}=-\frac{(-1)^{\Delta-1}}{\Gamma(\Delta)}\ln(1-t^{2})(\frac{\partial}{\partial s})^{\Delta-1}\int_{0}^{1}d\tau\frac{\tau^{p-1}}{\omega}+\ln(1-t^{2})o(1-t^{2})
\end{equation}

\begin{equation}
=-2^{\Delta-1}\ln(1-t^{2})\int_{0}^{1}d\tau\frac{\tau^{p-1}(1-\tau)^{\Delta-1}}{(2s+(1-2s)\tau)^{\Delta}}
\end{equation}

\begin{equation}
=-\frac{1}{2}\frac{\ln(1-t^{2})}{s^{\Delta}}\int_{0}^{1}d\tau\tau^{p-1}(1-\tau)^{\Delta-1}+o(1-2s)
\end{equation}
$\tau$ integral is Euler's beta function 

\begin{equation}
I_{\log}=-\frac{1}{2}\frac{\ln(1-t^{2})}{s^{\Delta}}\frac{\Gamma(p)\Gamma(\Delta)}{\Gamma(p+\Delta)} + o(1-2s).
\end{equation}
We obtain the s-channel leading logrithmic term in $D_{pp\Delta\Delta}$

\begin{equation}
D_{pp\Delta\Delta}|_{\log}=-\frac{\pi^{2}}{2}\frac{\Gamma(p+\Delta-2)}{\Gamma(p+\Delta)}\frac{1}{(x_{13}^{2})^{p}(x_{24}^{2})^{\Delta}}\ln(1-t^{2}),
\end{equation}
which is in agreement with leading logarithmic term in equations (6.30) and (A.3) of Ref. \cite{hfmmr}.

The u-channel limit  $(|x_{14}|\ll |x_{13}|, |x_{23}|\ll |x_{13}|)$ correspond to $s\rightarrow \frac{1}{2}$ and $t\rightarrow 1$ and is given by the same expression.
 
The full correlator $<O_{\phi}(\vec x_{1})O_{C}(\vec x_{2})O_{\phi}(\vec x_{3})O_{C}(\vec x_{4})>$ can be expressed in terms of quartic graphs of the form $D_{pp\Delta\Delta}$. We simply take the result from refs. Refs. \cite{agmoo,hfmmr}.
\begin{equation}
I=(\frac{6}{\pi^{2}})^{4}[16 x_{24}^{2} (\frac{1}{2s}-1) D_{4455} + \frac{64}{9}\frac{x_{24}^{2}}{x_{13}^{2}}\frac{1}{s} D_{3355} + \frac{16}{3}\frac{x_{24}^{2}}{x_{13}^{4}}\frac{1}{s} D_{2255}
\end{equation}
\[
- 14 D_{4444} - \frac{46}{9 x_{13}^{2}} D_{3344} - \frac{40}{9 x_{13}^{4}} D_{2244} - \frac{8}{3 x_{13}^{6}} D_{1144} + 64 x_{24}^{2}D_{4455}],
\]
where s is the conformally invariant variable as above.

Using the results for leading logarithmic singularity in direct and crossed channel limit of $D_{pp\Delta\Delta}$, we can write the result for the full correlator.\\ 
In direct channel ($x_{13}\ll x_{12}, x_{24}\ll x_{12}$) limit we obtain the leading logrithmic term as below
\begin{equation}
I_{\log}=\frac{16\times 8}{27\times 7}\frac{\pi^{2}}{2}(\frac{6}{\pi^{2}})^{4}\frac{1}{(x_{12}^{2})^{4}(x_{34}^{2})^{4}}\ln s,
\end{equation}
which is of the form
\[
I_{\log}\sim \frac{1}{(x_{12}^{2})^{4}(x_{34}^{2})^{4}}\ln(\frac{x_{13}x_{24}}{x_{12}x_{34}}).
\]
In s-channel  ($x_{12}\ll x_{14}, x_{34}\ll x_{14}$) limit we obtain the leading logarithmic term as below
\begin{equation}
I_{\log}=-\frac{8}{9}\frac{\pi^{2}}{2}(\frac{6}{\pi^{2}})^{4}\frac{\ln (1-t^{2})}{(x_{13}^{2})^{4}(x_{24}^{2})^{4}},
\end{equation}
which is of the form
\[
I_{\log}\sim \frac{1}{(x_{13}^{2})^{4}(x_{24}^{2})^{4}}\ln(\frac{x_{12}x_{34}}{x_{14}x_{23}}).
\]
Similarly in u-channel ($x_{14}\ll x_{12}, x_{23}\ll x_{12}$) limit the leading singularity is of the form
\[
I_{\log}\sim \frac{1}{(x_{13}^{2})^{4}(x_{24}^{2})^{4}}\ln(\frac{x_{14}x_{23}}{x_{12}x_{34}}).
\]

We conclude that leading logarithmic singularity in crossed channels is also due to an operator of dimension eight. Also we note that the co-efficient of leading singularity is same in both s- and u- channels as expected and different from that one in t-channel. (Direct channel (t-channel) involes $\phi\phi C C$ scattering and both crossed channels (s- and u-channels) involve $\phi C\phi C$ scattering where $O_{\phi}\sim Tr(F^{2} + ...), O_{C}\sim Tr(F\tilde{F} + ...)$). 

\vspace{5mm}
{\bf Acknowledgements:}
We are grateful to Ian Kogan for useful discussions and comments. This work is supported by Felix Scholarships.

\end{document}